\documentclass{article}
\usepackage{amsmath}
\usepackage{amssymb}

\input{tcilatex}
\begin{document}

\author{Richard Healey \\
University of Arizona}
\title{Quantum States as Objective Informational Bridges}
\date{February 1st, 2015}
\maketitle

\begin{abstract}
A quantum state represents neither properties of a physical system nor
anyone's knowledge of its properties. The important question is not what
quantum states represent but\ how they are used---as informational bridges.
Knowing about some physical situations (its backing conditions), an agent
may assign a quantum state to form expectations about other possible
physical situations (its advice conditions). Quantum states are objective:
only expectations based on correct state assignments are generally reliable.
If a quantum state represents anything, it is the objective probabilistic
relations between its backing conditions and its advice conditions. This
paper offers an account of quantum states and their function as
informational bridges, in quantum teleportation and elsewhere.
\end{abstract}

\newpage

\begin{quotation}
the quantum state of a Qbit or a collection of Qbits is not a property
carried by those Qbits, but a way of concisely summarizing everything we
know that has happened to them, to enable us to make statistical predictions
about the information we might then be able to extract from them. (N. David
Mermin [2007, p. 109])
\end{quotation}

\section{Introduction}

In his recent extended review article, Matt Leifer [2014] classifies the
pragmatist view of quantum theory sketched in my [2012a] as $\psi $%
-epistemic rather than $\psi $-ontic while acknowledging that the $\psi $%
-ontology theorems he discusses do not apply to it. But as he draws the $%
\psi $\textit{-ontic/}$\psi $\textit{-epistemic} distinction in his
introduction, that pragmatist view counts as neither $\psi $-epistemic nor $%
\psi $-ontic. It does not view $\left\vert \psi \right\rangle $\ as
describing some independently existing property of the system to which it is
assigned, but nor does it take a quantum state to exist only in the minds
and calculations of quantum theorists. This paper is intended to clarify my
[2012a] view of the quantum state, and to show how $\left\vert \psi
\right\rangle $ refers to something that objectively exists in the world,
independently of any observer or agent, and would still exist if all
intelligent beings were suddenly wiped out from the universe---which is how
Leifer characterizes an \textit{ontic} state.

\qquad A quantum state (given by a wave-function, vector or density
operator) represents neither properties of a physical system to which it is
assigned, nor anyone's knowledge of its properties. The important question
is not what quantum states represent but\ how they are used---as
informational bridges. Knowing about some physical situations (its backing
conditions), an agent may assign a quantum state to form expectations about
certain other possible physical situations (its advice conditions). Quantum
states are objective: only expectations based on correct state assignments
are generally reliable. But a quantum state represents neither its backing
conditions nor its advice conditions.

\qquad Since its main function is to provide information on what advice
conditions to expect given prevailing backing conditions, it may be said to
represent probabilistic relations between backing conditions and advice
conditions. These probabilistic relations are objective: they would exist in
a world without agents, as long as that world featured patterns of
statistical regularity that were sufficiently stable to be modeled by Born
probabilities of the values of physical magnitudes that specify what I have
called advice conditions. We can use quantum theory successfully because (or
at least in so far as) they \textit{do} exist in our world.

\qquad As physically situated, and so epistemically limited, a user of
quantum theory can assign quantum states on the basis of what that agent is
in a position to know, in order to form reasonable expectations about what
that agent is not in a position to know. That is how quantum states function
as informational bridges. A quantum state provides a sturdy informational
bridge only if it would be the right state to assign for any agent in that
physical, and \textit{therefore} epistemic, situation---only if it refers to
the \textit{actual} probabilistic relations between its accessible backing
conditions and inaccessible advice conditions.

\qquad Here is the structure of the rest of the paper. By distinguishing
between quantum mathematical models and their physical applications section
2 shows how a precise formulation of quantum theory may be given without
using terms like `measurement', `observation', `information' or `agent'. Any
account of how the theory is applied must mention agents who may apply it:
agents are (potential) \textit{users} of quantum models---they are not
mentioned in those models. Any application of a quantum model is \textit{%
perspectival}---it is from the perspective of a hypothetical, physically
situated, agent. So to say how a model of quantum theory is applied we need
a rough physical characterization of this situation as what I will call an 
\textit{agent-situation}. Section 2 supplies this and clarifies the notions
of backing conditions and advice conditions. A correct quantum state
assignment is relative to an agent-situation because these conditions are a
function of that situation. Section 3 shows the worth of these ideas in
explaining how quantum theory works in some applications that have been
thought to raise problems of "wave collapse" and causal anomalies. Chris
Timpson [2013] used his analysis of quantum information$_{t}$ as a sequence
of quantum states to demystify the issue of information flow in quantum
teleportation. But there is more to be said if a quantum state itself
functions as an informational bridge: I say it in section 4. The paper
concludes by summarizing its main points and relating them to recent work in
quantum foundations.

\section{Quantum states: Objective but relational}

In discussing quantum states we should start by distinguishing the state
itself from any of its mathematical representatives, including $\left\vert
\Psi \right\rangle $ (or a wave-function representation such as $\psi (%
\mathbf{r}_{1},...,\mathbf{r}_{n})$), $\hat{\rho}$, or (as in AQFT) $\omega :%
\mathcal{A}\longrightarrow 
\mathbb{C}
$. As a mathematical object, none of these representatives of a quantum
state exists in the physical world: what is at issue is the nature and
function of the quantum state such objects purport to represent. But that
issue is best approached indirectly, by studying the role of a mathematical
representative of a quantum state in an application of a mathematical model
of quantum theory of which it is one element.

\qquad Such models take different forms in different kinds of quantum
theory, but since the same issue arises for each I'll consider only models
of non-relativistic quantum mechanics of the form $\Theta \mathcal{=}%
\left\langle \mathcal{H},T,\Psi ,\mathcal{A},\mu \right\rangle $. In an
application, the interval of real numbers $T$ will represent an interval of
time while $t\in T$ represents a moment in that interval. Each operator in $%
\mathcal{A}$ is linear: in an application a self-adjoint operator $\hat{A}$
will correspond uniquely to a dynamical variable $A$. The self-adjoint
operator $\hat{H}$ corresponding to energy is associated with a family of
unitary operators $\hat{U}_{t}=\exp i\hat{H}t$ that define the possible
trajectories of $\left\vert \Psi (t)\right\rangle $ (in the Schr\"{o}dinger
picture). The measure $\mu $ assigns, at each $t\in T$, a number in the
interval [$0,1$] to each subspace $\mathcal{K}$\ of $\mathcal{H}$ such that $%
\mu (\mathcal{H})=1$, $\mu (\mathcal{\varnothing })=0$ and if $\mathcal{K}%
_{i}\perp $ $\mathcal{K}_{j}$ for $i\neq j$ then $\mu (\mathcal{L}%
)=\tsum\nolimits_{i}\mu (\mathcal{K}_{i})$, where $\mathcal{L}=\vee _{i}%
\mathcal{K}_{i}$ is the span of the $\mathcal{K}_{i}$. In an application 
\textit{some} values of $\mu $ may yield Born probabilities.

\qquad To apply a model of the form $\Theta $ one first chooses the
dimension of a Hilbert space $\mathcal{H}$ in which to represent a quantum
state $\Psi $ of a physical system $s$ over a period of time $T$ and a
particular $\hat{H}$ on $\mathcal{H}$: then one assigns a particular initial
state $\left\vert \Psi (t_{0})\right\rangle $ to $s$. This picks out a
unique trajectory $\left\vert \Psi (t)\right\rangle $ and so a particular
state $\left\vert \Psi (t_{1})\right\rangle $ at a later time $t_{1}\in T$.
Next one chooses a pairwise-commuting family $\mathcal{F}=\{\hat{A},\hat{B}%
,...\}$\ of self-adjoint operators on $\mathcal{H}$\ corresponding to
dynamical variables $A,B,...$ respectively. One can now apply the Born rule
in the form%
\begin{gather}
\Pr ([A\in \Gamma ]\&[B\in \Delta ]\&...)=\left\langle \Psi
(t_{1})|P^{A}(\Gamma ).P^{B}(\Delta )...|\Psi (t_{1})\right\rangle \\
=\mu _{\Psi (t_{1})}(\mathcal{M}_{\Gamma }^{A}\wedge \mathcal{M}_{\Delta
}^{B}\wedge ...)  \notag
\end{gather}%
where the elements $P^{A}(\Gamma )$, $P^{B}(\Delta ),..$ of the spectral
measures of $\hat{A},\hat{B},...$\ project onto subspaces $\mathcal{M}%
_{\Gamma }^{A},\mathcal{M}_{\Delta }^{B},...$ respectively of $\mathcal{H}$.
This yields a joint probability distribution for the values of dynamical
variables $A,B,...$ on $s$ at $t_{1}$.

\qquad While nowhere explicitly mentioning measurement or preparation, this
familiar account may seem implicitly to depend on these notions. Doesn't
assignment of a particular state to a system at $t_{0}$\ require knowledge
of how this has been prepared? Don't Born probabilities refer only to the
outcome of a joint measurement at $t_{1}$\ of a compatible family of
dynamical variables defined by the measurement context? And doesn't the
applicability of the model at time $t_{1}$ presuppose that no measurement
occurred between $t_{0}$ and $t_{1}$?

\qquad In fact one can say what warrants assignment of a quantum state and
what circumstances license application of the Born rule without mentioning
preparation or measurement. In each case it is simply the prevailing
physical conditions that supply the necessary warrant, whether or not anyone
seeks to exploit these to prepare the system's state or make a measurement
on it. In the first case I call these backing conditions, since when they
obtain they back assignment of a particular quantum state. In the second
case environmental conditions determine when and how the Born rule may
legitimately be applied to provide reliable advice about the values of
certain dynamical variables, in the form of what I call advice conditions.

\qquad I headed this paper with a quote from David Mermin in which he
characterizes a quantum state of a system as a way of concisely summarizing
everything we know that has happened to it.\footnote{%
In two earlier papers (Mermin [2002], Brun, T., Finkelstein, J. and Mermin,
N.D. [2002]) he had raised and answered the question "Whose knowledge?".}
The knowledge of which he speaks is knowledge of backing conditions. Knowing
a state's backing conditions, one is justified in assigning that state: but
one would be \textit{warranted} in assigning the state whether or not one
knew these conditions, just as a test result may warrant a diagnosis whether
or not the doctor knows about it. I'll give several examples of backing
conditions later. For now it is important only that they may be stated in
purely physical language, with no talk of observers or agents and their
measuring or preparing activities. But nor does this statement involve talk
of quantum states, operators or Hilbert space measures: such talk is
confined to the mathematical model being applied.

\qquad Advice conditions are stated in a more restricted physical language
because of how they figure in the Born rule. When legitimately applied, that
rule yields probabilities for what I'll call \textit{magnitude claims}: 
\textit{The value of dynamical variable }$M$ \textit{on physical system }$s$%
\textit{\ lies in }$\Delta $ is a canonical magnitude claim which I'll write
as $M_{s}\in \Delta $. Magnitude claims are used to state advice
conditions---so-called because they are the topic of the Born rule's
probabilistic advice. But that advice is necessarily selective. The Born
rule yields joint probabilities only for magnitude claims corresponding to 
\textit{commuting} projection operators, and Gleason's [1957] theorem and
other results show that these typically cannot all be recovered as marginals
of any underlying joint distribution.

\qquad Restrictions on use of the Born rule are needed to ensure its
consistent application to a system, but these arise naturally from the
physical circumstances in which that system finds itself. $\Theta $\
provides an adequate\ quantum model of a system $s$\ only while $s$ remains
isolated from interaction with other quantum systems: If $s$ is subjected to
such interactions, this can also be modeled by the unitary evolution of a
joint quantum state in an expanded Hilbert space. In various models of
decoherence this extremely rapidly and essentially irreversibly results in
the delocalization of phase of the quantum state of $s$ and/or a system $%
\alpha $ with which it interacts into the wider environment, thereby picking
out a preferred "pointer basis" of orthogonal subspaces in their Hilbert
space(s).

\qquad The Born rule may be legitimately applied to $s$ only when such
decoherence has occurred, and then only to those privileged magnitude claims
corresponding to projection operators onto subspaces in the relevant pointer
basis.\footnote{%
For further details, see my [2012b].} Such decoherence is never perfect, and
nor is a "pointer basis" precisely determined and perfectly constant. But
the advice provided by the Born rule concerning only magnitude claims
privileged by pointer bases in the same narrow neighborhood will be
consistent and typically prove reliable: the corresponding advice conditions
will typically obtain with relative frequencies closely corresponding to
their Born rule probabilities.

\qquad You and I may know different things about what has happened to a
quantum system. Should we then assign the system different quantum states?
If you have simply noticed something I missed, or I have forgotten something
you remember, then one of us has not made use of all the available
information: our quantum state assignment should be based on all the
information to which we have access. But the information to which an agent
has access depends on that agent's physical situation.

\qquad A single agent gains access to additional information merely with the
passage of time, while spatially separated agents have access to different
information just because they are in different places. It is not only
spatiotemporal location that imposes physical limits on one's knowledge of
what has happened to a system. Acquisition of observational knowledge
depends on the presence of physical channels capable of conveying that
knowledge to the observer, which is why it is so hard to observe dark matter
in distant galaxies. The upshot is that when an agent assigns a quantum
state the conditions backing that assignment are a function of the agent's
physical situation.

\qquad We often apply quantum theory not to the actual world but to some
simplified or otherwise modified variant of it---to a merely possible world.
Clearly the agent applying the model is not located in that merely possible
world. But even such applications are from the perspective of some merely
possible agent in that world---the perspective of what I will call an 
\textit{agent-situation}. All applications of quantum theory are from the
perspective of a physically-characterized agent-situation. Applications to
merely possible worlds make it particularly clear that no agent need occupy
such an agent-situation.

\qquad An agent-situation is characterized, in part, by a space-time region
marking the momentary location of a hypothetical agent. It is common to
idealize the histories of observers by timelike curves in a relativistic
space-time---their world-lines. In this idealization the space-time location
of an agent-situation would be a point $p$\ on such a curve. Since no
physical processes can convey information faster than light, no agent at $p$
would have observational access to anything outside the past light cone of $p
$. So the backing conditions for a quantum state assignment relative to $p$
lie in (or on) the past light cone of $p$.

\qquad A quantum state assignment to a system relative to $p$ will be a
function of everything in the past light cone of $p$ (though much of what
happens there will prove irrelevant). But it is important to notice that
quantum state assignments relative to $p$,$q$ may differ: conditions
obtaining in the past light cone of $p$ may back a different assignment than
conditions obtaining in the past light cone of $q$. Since every quantum
state assignment is relative to an agent-situation it is misleading to speak
of a system's being in a quantum state, as if this expressed a property of
that system rather than a relation\ to the agent-situation from whose
perspective it is assigned.

\section{Some applications}

Before moving on to its application to the phenomenon of quantum
teleportation, I'll consider two illustrative examples of this account of
quantum states as objective informational bridges---violation of Bell
inequalities and delayed choice entanglement-swapping.

\qquad Quantum theory correctly predicts the violations of Bell inequalities
that have been observed in experiments involving space-like separated
measurements of linear polarization on polarization-entangled photon pairs
(EPR-Bohm pairs) (Weihs \textit{et al}. [1998]). Bell [2004] used these
predictions to argue that quantum theory is not a locally causal theory.
Maudlin [2011] argued further that space-like (superluminal) influences must
be present in experiments verifying them: Price [1996], on the other hand,
suggested their results might be explained by retrocausal influences. One
way for a measurement on one photon in an EPR-Bohm pair to influence the
result of a spacelike separated measurement on the other photon would be
through "collapsing its quantum state"---just the kind of "spooky" action at
a distance Einstein rejected. Any such instantaneous collapse would threaten
to reintroduce an absolute notion of simultaneity into a relativistic
space-time from which it had been successfully excised.

\qquad In fact quantum theory may be applied to explain EPR-Bohm
correlations with no superluminal or retro- causation and no physical
quantum state reduction.\footnote{%
For a fuller account see my [forthcoming].} The relativization of quantum
states to agent-situation is key to this explanation. An entangled state $%
\Phi ^{+}=1/\surd 2(\left\vert HH\right\rangle +\left\vert VV\right\rangle )$
may be correctly assigned to the pair relative to an agent-situation with
location $p$\ in the past light cone of either of the space-like separated
polarization measurement events: its backing conditions (in the past light
cone of $p$) include whatever physical conditions the experimenters used to
prepare that state, such as parametric down-conversion of laser light by
passage through a non-linear crystal.

\qquad Because of decoherence at the detectors, the Born rule is
legitimately applied to this state to yield an equal chance of either
outcome of any linear polarization measurement at either detector, but
chance $\cos ^{2}\angle ab$ that the two detectors will yield the same
outcome for linear polarization measurements with respect to axes inclined
at an angle $\angle ab$ to one another. It is important to stress that these
chances are \textit{also} relative to agent-situation with location $p$. For
anyone who accepts quantum theory, they give the objectively correct advice
to one located at $p$ about how firmly to believe the corresponding
outcome(s) will be recorded. Thus state $\Phi ^{+}$ acts as an informational
bridge between its backing conditions and its advice conditions by enabling
anyone informed of its backing conditions to form reliable expectations
concerning it advice conditions.

\qquad Consider instead an agent situation with location $q$ in the future
light cone of recording event $\mathbf{1}$ but not of recording event $%
\mathbf{2}$. Relative to $q$, the correct quantum polarization state to
assign to the distant photon in the past light cone of $\mathbf{2}$ depends
on the outcome at $\mathbf{1}$: suppose it is $V_{a}$, where linear
polarization at $\mathbf{1}$ was measured with respect to axis $a$. Then $%
\left\vert V_{a}\right\rangle $ is the correct quantum state to assign to
the distant photon, relative to agent situation with location $q$. This is
not because "the" quantum state of the pair has collapsed: the correct
quantum state to assign relative to agent-situation with location $p$ is
still $%
{\frac12}%
\hat{1}$---the reduced state of $\Phi ^{+}$. The correct state relative to
agent situation with location $q$ is $\left\vert V_{a}\right\rangle $
because the outcome at $\mathbf{1}$ is in the past light cone of $q$, and so
counts as an additional accessible backing condition determining the
assignment of a quantum state relative to $q$. Relative to agent situation
with location $q$ the chance of outcome $V_{b}$ at $\mathbf{2}$ is $\cos
^{2}\angle ab$, as follows from the legitimate application of the Born rule
to state $\left\vert V_{a}\right\rangle $. Here it is state $\left\vert
V_{a}\right\rangle $ that acts as an informational bridge between its
(augmented) backing conditions and its advice conditions.

\qquad These applications of quantum theory explain the patterns of
correlation that are taken to violate CHSH inequalities by showing that they
were to be expected and what they depend on. Both individual outcome events
and the event of their joint occurrence depend causally on the physical
conditions backing assignment of $\Phi ^{+}$ since an agent could affect the
chances of these events by modifying those backing conditions. But there is
no causal dependence of one outcome event on the other, since no-one who
accepts quantum theory can countenance the possibility of an agent's
modifying either outcome while keeping fixed both detectors' settings and
the conditions backing $\Phi ^{+}$. Nor do the detector settings have any
retrocausal influence on events acknowledged by quantum theory.

\qquad Now consider delayed-choice entanglement-swapping. Suppose two
EPR-Bohm photon pairs $\langle 1,2\rangle $, $\langle 3,4\rangle $\ are
independently prepared in conditions backing assignment of Bell states $\Psi
_{12}^{-},\Psi _{34}^{-}$%
\begin{equation}
\Psi ^{-}=1/\surd 2(\left\vert HV\right\rangle -\left\vert VH\right\rangle ).
\end{equation}
First assume a Bell-state measurement is conducted on $\langle 2,3\rangle $,
and the result noted. Conditional on a particular result, one of four
different entangled states may then be assigned to the pair $\langle
1,4\rangle $. This is entanglement-swapping.\footnote{%
By adding a classical channel it may be used to teleport entanglement of a
Qbit.} The particular entangled state of $\langle 1,4\rangle $ may be
verified by quantum tomography in the usual way, and violation of Bell
inequalities demonstrated. Alternatively, assume the linear polarizations of
each of $2,3$ are measured independently (instead of the Bell-state
measurement). No matter what the outcomes of the measurements on $2,3$,
measurements of linear polarization on $1,4$\ will reveal no correlations:
since the states then assigned to $1,4$\ are not entangled their joint
probability distribution will factorize.

\qquad So far I \ have said nothing about when and where the measurements on 
$1,4$ occur, relative to those on $2,3$. In a standard case of
entanglement-swapping, the Bell-state measurement $B_{23}$ on $2,3$ occurs
time-like earlier than each of the measurements $M_{1},M_{4}$ on $1,4$. But
the correlations between the outcomes of measurements $M_{1},M_{4}$\
conditional on a particular (joint) outcome of a Bell-state measurement $%
B_{23}$ do not depend on the space-time intervals $I(B_{23},M_{1})$, $%
I(B_{23},M_{4})$. If $B_{23}$ occurs timelike \textit{later} than $%
M_{1},M_{4}$ then we have an apparent case of "retrocausal entanglement
swapping". Since the choice of whether to perform $B_{23}$ or independent
polarization measurements $M_{2},M_{3}$ may be made at random \textit{after}
photons $1,4$ have been absorbed into their detectors, this has also been
called delayed choice entanglement-swapping.\footnote{%
By Peres [2000]. This has been realized experimentally by Ma \textit{et al. }%
[2013].}

\qquad Relative to an agent-situation located just in the overlap of the
future light cones of their production events, the correct initial
polarization state $\Psi $\ to assign to the photons is%
\begin{equation}
\left\vert \Psi _{12}^{-}\right\rangle \left\vert \Psi
_{34}^{-}\right\rangle =1/2(\left\vert \Psi _{14}^{+}\right\rangle
\left\vert \Psi _{23}^{+}\right\rangle -\left\vert \Psi
_{14}^{-}\right\rangle \left\vert \Psi _{23}^{-}\right\rangle -\left\vert
\Phi _{14}^{+}\right\rangle \left\vert \Phi _{23}^{+}\right\rangle
+\left\vert \Phi _{14}^{-}\right\rangle \left\vert \Phi
_{23}^{-}\right\rangle .  \label{4-fold entangled state}
\end{equation}%
Analysis of the actual experimental setup of Ma \textit{et al.} [2013] shows
that cases (i) in which $B_{23}$ records one photon as detected to each side
of the beam splitter (with the same polarization) have non-zero Born
probability only from the fourth term in (\ref{4-fold entangled state}),
while cases (ii) in which $B_{23}$ records both photons as detected to the
same side of the beam splitter (with opposite polarizations) have non zero
Born probability only from the third term in (\ref{4-fold entangled state}).
What \ would be the correct state to assign to the remaining photons,
relative to an agent-situation located just in the future light cone of $%
B_{23}$? In case (i) the corresponding $\langle 1,4\rangle $ pair should be
assigned the quantum state $\Phi _{14}^{-}$, while in case (ii) the
corresponding pair should be assigned $\Phi _{14}^{+}$. These would be the
correct state assignments irrespective of the space-time intervals $%
I(B_{23},M_{1})$, $I(B_{23},M_{4})$.

\qquad A\ recent paper (Egg [2013, p.1133]) rejects that conclusion as
follows (I have changed his notation to conform to mine):

\begin{quotation}
The Bell measurement on the $\langle 2,3\rangle $ pair allows us to sort the 
$\langle 1,4\rangle $ pairs\ into four subensembles corresponding to the
four Bell states. Without delayed choice, this has physical significance,
because each $\langle 1,4\rangle $ pair \textit{really is} in such a state
after the $\langle 2,3\rangle $ measurement [$B_{23}$]. But if the $\langle
1,4\rangle $ measurements [$M_{1},M_{4}$] precede [$B_{23}$], the $\langle
1,4\rangle $ pair \textit{never is in any of these states}. ... Therefore,
far from being committed\ to any indeterminacy about entanglement (or any
backward-in-time influences) a realistic view of the quantum state yields a
perfectly clear assessment of what happens in entanglement-swapping: If [$%
B_{23}$] occurs at a time the complete system is still in state $\Psi $, it
confers entanglement on the $\langle 1,4\rangle $ pair, if it occurs at a
later time it does not.
\end{quotation}

\qquad In section 2 I warned against talk of a system's being in a quantum
state (as opposed to being assigned the correct quantum state relative to a
particular agent-situation). We can see how this leads to problems by
reflecting on a case of entanglement-swapping in which the intervals $%
I(B_{23},M_{1})$, $I(B_{23},M_{4})$ are space-like. In this case the
distinction to which Egg appeals has no invariant significance: whether $%
M_{1}$ or $M_{4}$ precedes $B_{23}$ depends on an arbitrary choice of
reference frame. Egg is not unaware of this problem: he admits (p.1130) that
his "reality-of-states" view should take \textit{state reduction}\
seriously, so that the change of description on "measurement" corresponds to
a real physical change. His "solution" ( footnote 7, p.1130) is to appeal to
a definite (although undetectable) temporal ordering between any two
events---a preferred foliation of space-time such as that required by
alternative theories including Bohmian mechanics.

\qquad But there is no physical process of state reduction, and no need to
appeal to any undetectable space-time foliation to understand
entanglement-swapping or violation of Bell inequalities. The reassignment of
a quantum state associated with relativizing it to a different
agent-situation is not a physical process. It is typically required because
additional backing conditions are accessible from the second
agent-situation. The discussion of EPR-Bohm correlations provided an
example: $\left\vert V_{a}\right\rangle $ is the correct quantum state to
assign to the distant photon relative to agent-situation $q$ while the
correct quantum state to assign relative to agent-situation with location $p$
is $%
{\frac12}%
\hat{1}$: each of these quantum states is objective, but there is no
question as to which quantum state the distant photon "is in".

\qquad When an entangled state is assigned to the pair $\langle 1,4\rangle $
conditional on the outcome of $B_{23}$ this state functions as an objective
informational bridge in the usual way, irrespective of the space-time
intervals $I(B_{23},M_{1})$, $I(B_{23},M_{4})$. What distinguishes the case
in which $B_{23}$ occurs invariantly later than $M_{1},M_{4}$ is the fact
that \textit{if appropriate physical channels were in place} the outcomes of 
$M_{1},M_{4}$ would be accessible from the space-time location of $B_{23}$.
In that case there would be no point in using the entangled state assigned
to the pair $\langle 1,4\rangle $ as an informational bridge to form
expectations as to these outcomes, since knowledge as to what they were
would already be available at the space-time location of $B_{23}$.

\qquad This illustrates the important point that there is more to an
agent-situation than its space-time location. A physical characterization of
an agent situation located just in the future light cone of $B_{23}$
specifies either the presence or the absence of physical channels connecting
this location to the outcomes of $M_{1},M_{4}$: only if no such channels are
in place is it necessary to assign an entangled state to $\langle 1,4\rangle 
$ conditional on the outcome of $B_{23}$, since only relative to that
agent-situation does this state assignment provide the best available
information concerning the outcomes of $M_{1},M_{4}$. In a case in which $%
B_{23}$ occurs invariantly earlier than $M_{1},M_{4}$, or space-like
separated from them, an entangled state must always be assigned to $\langle
1,4\rangle $ relative to any agent-situation located just in the future
light cone of $B_{23}$ and backed by the outcome of $B_{23}$, because no
physical channels could give access to the outcomes of $M_{1},M_{4}$ from
that space-time location.

\qquad

\section{Information in quantum teleportation}

Chris Timpson devotes two chapters of his [2013] to a discussion of
information flow in quantum teleportation. His conclusion is largely
deflationary

\begin{quote}
we have been able to replace the (needlessly) conceptually puzzling question
of how the information$_{t}$ gets from Alice to Bob in teleportation, with
the simple, genuine question of what the physical processes involved in
teleportation are. (p. 86)
\end{quote}

He had previously proposed a helpful analysis of the technical notion of
information$_{t}$ after distinguishing this from our everyday, informal
notion. This locates quantum information$_{t}$ as a species of Shannon
information:

\begin{quote}
pieces of quantum information$_{t}$ are certain \textit{abstract}
items---sequences of quantum states (types)

As \textit{abstracta}, pieces of quantum information$_{t}$ ... do not
themselves have a spatio-temporal location; it is their tokens (if any)
which do. (p. 65)
\end{quote}

A token of quantum information$_{t}$ is then a concrete instance of an
abstract sequence of quantum states. The simplest example of a token of
quantum information$_{t}$ is a single (physical) Qbit. Quantum teleportation
is a physical process in which a Qbit associated with state $\left\vert \chi
\right\rangle $ is initially located only at $A$ and a Qbit associated with
state $\left\vert \chi \right\rangle $ is finally located at $B$ while the
Qbit formerly located at $A$ is no longer associated with state $\left\vert
\chi \right\rangle $. In this way the abstract piece of quantum information$%
_{t}$ previously instantiated only at $A$ is now instantiated only at $B$.

\qquad Among the remarkable features of quantum teleportation is that it
enables Alice to teleport an \textit{unknown} quantum state to Bob. After
its successful execution Alice and Bob both know that the state now
associated with Bob's Qbit is the same state formerly associated with
Alice's, but neither Alice nor Bob need know what this state is. A
teleported state may be unknown to Alice and Bob because neither of them
knows what had happened to Alice's Qbit before she got hold of it.

\qquad One can accept Timpson's helpful analysis without saying anything
about the nature and function of quantum states. But suppose the quantum
state of a Qbit is a way of concisely summarizing everything we know that
has happened to it. If there is no-one but Alice and Bob, then the quantum
state they teleport may be a way of concisely summarizing knowledge that
no-one knows---an unknown known! Clearly something has gone wrong. Not even
Donald Rumsfeld countenances unknown knowns.\footnote{%
Although the authors of a documentary film about him called it \textit{The
Unknown Known}, suggesting that he might have \textit{been} one.} Correcting
the errors leading to this paradoxical conclusion will prove very helpful in
arriving at a clearer understanding of quantum states as objective
informational bridges.

\qquad One can think of quantum teleportation as bridging an informational
gap by reproducing a token of quantum information$_{t}$ at a distant
location. But the notion of information involved in calling a quantum state
an informational bridge is not the technical notion (information$_{t}$)
analyzed by Timpson but the ordinary notion. A token of quantum information$%
_{t}$ is itself an informational bridge in this ordinary sense: it bridges
the gap between the state's backing conditions and its advice conditions. It
is backing conditions in the past light cone of $p$ that determine the
objectively correct initial quantum state to assign to Alice's Qbit,
relative to $p$. What those backing conditions are depends both on the
space-time location of $p$ and on the presence of physical channels capable
of giving access to them from $p$.

\qquad Consider Alice's situation at space-time location $A$ as she receives
her Qbit to be teleported to Bob at $B$. Reliable teleportation requires a
(classical) physical channel connecting $A$ to $B$. So whatever conditions $%
K $\ back assignment of initial state $\left\vert \chi \right\rangle $ to
Alice's Qbit relative to an agent-situation with location $A$ apparently
also back assignment of initial state $\left\vert \chi \right\rangle $ to
Alice's Qbit relative to an agent-situation with location $B$. If no
teleportation actually occurred, Bob would not have his own Qbit in state $%
\left\vert \chi \right\rangle $ at $B$, but he would appear to be in as good
a position as Alice to assign state $\left\vert \chi \right\rangle $ to
Alice's Qbit at $A$.

\qquad How could Alice \textit{not} know what state to assign to her Qbit at 
$A$? Perhaps Carol agreed to prepare this Qbit for Alice at $A$ by taking a
Qbit (they both agree to be) correctly assigned state $\left\vert
0\right\rangle $ and subjecting it to a unitary transformation into state $%
\left\vert \chi \right\rangle =a\left\vert 0\right\rangle +b\left\vert
1\right\rangle $ without telling Alice what that transformation was.
Whatever physical conditions $K^{\prime }$ involved in implementing Carol's
unitary transformation backing her assignment of $\left\vert \chi
\right\rangle $ occurred in the past light cone of $A$, but in the absence
of physical channels capable of giving an agent at $A$ access to $K^{\prime }
$\ those conditions would not back assignment of $\left\vert \chi
\right\rangle $ relative to Alice's agent-situation at $A$.\footnote{%
But doesn't Carol herself provide such a physical channel? Not if all traces
of what she did have been erased from her brain and the rest of her body.}
Then Alice at $A$ should not assign state $\left\vert \chi \right\rangle $
relative to her agent-situation, while correctly maintaining that her Qbit
should be assigned \textit{some} initial pure state, relative to a different
agent-situation such as Carol's just after implementing her unitary
transformation.

\qquad In fact Bob at $B$ is in a similar position. The existence of the
specific physical channel required for reliable teleportation does not
guarantee the existence of any other physical channel capable of giving
access at $B$\ to conditions $K$ that back assignment of initial state $%
\left\vert \chi \right\rangle $ to Alice's Qbit relative to her
agent-situation at $A$. If those conditions are physically inaccessible from 
$B$ then they do not back assignment of initial state $\left\vert \chi
\right\rangle $ to Alice's Qbit relative to Bob's agent-situation at $B$. In
that case Bob at $B$ should not assign a state to Alice's Qbit, while
correctly maintaining that her Qbit should be assigned \textit{some} initial
pure state, relative to a different agent-situation such as Alice's or
Carol's.

\qquad There is even a scenario in which initial pure state $\left\vert \chi
\right\rangle $ correctly assigned to Alice's Qbit by Carol (relative to her
agent-situation just after preparing it) is unknown to Alice but \textit{%
known} to Bob. This would involve the presence of physical channels
rendering the conditions $K^{\prime }$\ accessible from Bob's
agent-situation at $B$ but the absence of physical channels rendering
conditions $K^{\prime }$\ accessible from Alice's agent-situation at $A$. In
this scenario a quantum state unknown to Alice could be known to Bob whether
or not it was teleported from Alice to Bob!

\qquad To sum up, conditions in the past light cone of an agent-situation $G$%
\ may back assignment of a quantum state to a system relative to some other
agent-situation $H$, without backing assignment of a quantum state relative
to $G$. Alice and/or Bob may each be in such an agent-situation\ in a case
of quantum teleportation. That is how it can make sense to speak of
teleporting an unknown quantum state: it makes sense when the state
correctly assigned relative to Carol's agent-situation is unknown to Alice
and/or Bob, at least one of whom is in an agent-situation relative to which
no quantum state assignment is possible. Since an agent-situation need not
be occupied by any actual agent, Carol's presence is not required.

\section{Conclusion}

A quantum state functions as an informational bridge between its backing
conditions and its advice conditions. Since it is assigned from the
perspective of a physically-characterized agent-situation, what quantum
state is assigned is relative to that agent-situation. But this does not
make quantum state assignments subjective or dependent on the epistemic
state of any agent. Quantum state assignments are objectively correct or
incorrect: only correct state assignments are reliable informational
bridges. What makes them reliable are the patterns of statistical
correlation between their backing conditions and their advice conditions
that obtain in the physical world. There are objectively correct quantum
state assignments whether or not there are any agents capable of making
them. But it is physically situated, and so epistemically limited, agents
who naturally take the perspective of an agent-situation and so find it
useful to assign quantum states from that perspective.

\qquad Are quantum states\ real? If they are, what do they represent? There
are real patterns of statistical correlation in the physical world.
Correctly assigned quantum states reliably track these through legitimate
applications of the Born rule. But the Born rule does not directly\textit{\
describe} these patterns: the frequencies they display are not in exact
conformity to that rule---the unexpected does happen. A quantum state is
real in so far as it specifies the real probabilistic relations between its
backing conditions and its advice conditions. Those probabilistic relations
are objective. What they are depends on conditions in the physical world,
even though probability statements do not directly describe those
conditions. So a quantum state is not an element of physical reality (it is
not a beable): $\left\vert \psi \right\rangle $ does not represent a
physical system or its properties. But it is real nonetheless. The world
contains quantum states just as it contains probabilities, physicists and
their\ journals. 

\end{document}